\documentclass[preprint]{aastex}

\newcommand {\asec} {$^{\prime\prime}$}

\def\amin{\ifmmode ^{\prime}\else$^{\prime}$\fi}
\def\asec{\ifmmode ^{\prime\prime}\else$^{\prime\prime}$\fi}

\def\etal{{\it et\,al.\,}}

\begin{document}
%

%
\title{The European Large Area ISO Survey V: a {\it BeppoSAX} hard X-ray survey of the S1 region}
%

\author{D.M.~Alexander\altaffilmark{1,2},
F.~La Franca\altaffilmark{3},
F.~Fiore\altaffilmark{4},
X.~Barcons\altaffilmark{5},
P.~Ciliegi\altaffilmark{6},
L.~Danese\altaffilmark{1},
R.~Della Ceca\altaffilmark{7},
A.~Franceschini\altaffilmark{8},
C.~Gruppioni\altaffilmark{8},
G.~Matt\altaffilmark{3},
I.~Matute\altaffilmark{3},
S.~Oliver\altaffilmark{9},
F.~Pompilio\altaffilmark{1},
A.~Wolter\altaffilmark{7},
A.~Efstathiou\altaffilmark{10},
P.~Heraudeau\altaffilmark{11},
G.C.~Perola\altaffilmark{3},
M.~Perri\altaffilmark{12},
D.~Rigopoulou\altaffilmark{13},
M.~Rowan-Robinson\altaffilmark{10}
and S.~Serjeant\altaffilmark{10}}

\affil{$^1$International School for Advanced Studies, SISSA, via Beirut 2-4, I-34014,
Trieste, Italy}
\affil{$^2$Present address: Department of Astronomy and Astrophysics, 525 Davey Laboratory, Pennsylvania State University, University Park, PA 16802\\
email: davo@astro.psu.edu}
\affil{$^3$Dipartimento di Fisica, Universit\`a degli Studi ``Roma Tre'', via Della
Vasca Navale 84, I-00146, Roma, Italy}
\affil{$^4$Osservatorio Astronomico di Roma, via Frascati 33, Monteporzio,
Italy}
\affil{$^5$Instituto de F\'\i sica de Cantabria (CSIC-UC), 39005 Santander, Spain}
\affil{$^6$Osservatorio Astronomico di Bologna, via Ranzani 1, Bologna, Italy}
\affil{$^7$Osservatorio Astronomico di Brera, via Brera 28, Milano, Italy}
\affil{$^8$Osservatorio Astronomico di Padova, Vicolo dell'Osservatorio 5,
Padova, Italy}
\affil{$^9$Astronomy Centre, CPES, University of Sussex, Falmer, Brighton, UK}
\affil{$^{10}$Astrophysics Group, Imperial College of
Science Technology \& Medicine, Prince Consort
Road, London, UK}
\affil{$^{11}$Max-Planck-Institut fuer Astronomie (MPIA), Koenigstuhl 17, D-69117 Heidelberg, Germany}
\affil{$^{12}${\it BeppoSAX} Science Data Center, ASI, Via Corcolle 19, Roma, Italy}
\affil{$^{13}$Max-Planck-Institut fur extraterrestrische Physik (MPE),
P.O.Box 1603, 85740 Garching, Germany}

\vfill\eject
%
\begin{abstract}
%

We present {\it BeppoSAX} observations of the Southern S1 region in
the European Large Area {\it ISO} Survey (ELAIS). These observations
cover an area of $\sim$1.7 sq. deg. and reach an on-axis ($\sim$0.7
sq. deg) 2-10 keV (HX) sensitivity of $\sim$10$^{-13}$ ${\rm erg}\, {\rm
s}^{-1}\, {\rm cm}^{-2}$. This is the first HX analysis
of an {\it ISOCAM} survey. We detect 9 sources with a signal to noise ratio
SNR$_{HX}>$3, 4 additional sources with a 1.3 to 10 keV (T) SNR$_T>$3 and 2 additional sources which appear to be associated with QSOs with SNR$_T>$2.9. The number densities of the SNR$_{HX}>$3 sources are consistent with the {\it ASCA} and
{\it BeppoSAX} logN-logS functions. Six {\it BeppoSAX} sources have reliable
{\it ISOCAM} 15$\mu$m counterparts within $\sim$60
arcsec. All these {\it ISOCAM} sources have optical counterparts of R$<$20 mags. Five of these sources have been previously optically classified giving 4
QSOs and 1 BALQSO at $z$=2.2. The remaining unclassified source has X-ray and photometric properties consistent with that of a nearby Seyfert galaxy. One further HX source has a 2.6$\sigma$ {\it ISOCAM} counterpart associated with a galaxy at z=0.325. If this {\it ISOCAM} source is real, the HX/MIR properties suggest either an unusual QSO or a cD cluster galaxy. We have
constructed MIR and HX spectral energy distributions to compute the
expected HX/MIR ratios for these classes of objects up to $z$=3.2 and
assess the HX/MIR survey depth.

The BALQSO has an observed X-ray softness ratio and HX/MIR flux ratio
similar to QSOs but different to those found for low redshift
BALQSOs. This difference can be explained in terms of absorption, and
suggests that high redshift BALQSOs should be comparatively easy to
detect in the HX band, allowing their true fraction in the high redshift QSO
population to be determined.

The QSOs cover a wide redshift range (0.4$<z<$2.6) and have HX/MIR flux ratios
consistent with those found for nearby {\it IRAS} and optically selected PG
QSOs. This suggests that MIR selected QSOs of R$<$20 mags come from the same
population as optically selected QSOs. We confirm this with a
comparison of the B/MIR flux ratios of MIR and blue band selected QSOs.

\end{abstract}


%
\section{INTRODUCTION}
%

Unified models of Active Galactic Nuclei (AGN) propose that all types
of AGN are fundamentally the same although the presence of a dusty
molecular torus prevents the observation of the broad line region for
particular line of sights. In this picture Type 1 AGN (e.g. Seyfert
1/QSO) are those for which the nucleus can be directly observed, while
in Type 2 AGN (e.g.  Seyfert 2) it is obscured by the torus (e.g.\
Lawrence and Elvis, 1982 and Antonucci, 1993). Strong support for this
model has been given by X-ray, near-IR and mid-IR observations
showing that Type 2 AGN are characterized by strong absorption whilst
Type 1 AGN are relatively unabsorbed (Turner \etal 1997,
Alonso-Herrero, Ward, Kotilainen, 1997, Maiolino \etal 1998 and
Clavel \etal 2000). In this framework the infrared radiation is
interpreted in terms of thermal emission from hot dust grains heated
by the high energy central source emission (optical to X-ray
continuum, see e.g. Granato, Danese, Franceschini 1997).

Barcons \etal (1995 hereafter B95) investigated the 12$\mu$m to hard X-ray
(HX, 2 to 10 keV) flux ratios of a sample of nearby AGN and galaxies
using {\it IRAS} and {\it HEAO1 A-2} observations. They found that Type 1 AGN have
larger HX/12$\mu$m flux ratios than Type 2 AGN whilst normal galaxies have
very weak HX emission. This implies high column densities
(log(N$_{H}$)$>$22 cm$^{-2}$) in the Type 2 objects, supporting the
basic unified model hypothesis. The study of B95 was necessarily
restricted to nearby objects. The investigation of this flux ratio to
deeper fluxes and higher redshifts will probe the properties of
luminous sources such as QSOs and Broad Absorption Line QSOs (BALQSOs)
and provide clues to the constancy of AGN activity in the universe.

The {\it ASCA} and {\it BeppoSAX} X-ray satellites have been
successful in determining the HX properties of AGN and have resolved
$\sim$30\% of the HX background into discrete sources, mostly Type 1
and Type 2 AGN (e.g.\ Ueda \etal 1998, 1999, Cagnoni \etal 1998,
Fiore \etal 1999 and Akiyama \etal 2000). The recent launch of the
{\it Chandra} and {\it XMM-Newton} X-ray satellites have now made it
possible to probe to fainter fluxes at HX energies. Recent {\it Chandra}
observations (e.g.\ Mushotzky \etal 2000, Fiore \etal 2000a, Brandt \etal
2000, Giacconi \etal 2000 and Hornschemeier \etal
2000) have uncovered a number of HX emitting, but optically apparently
normal galaxies. The HX slopes of the sources are sufficiently hard
that, when combined with the HX emitting AGN sources, they can account
for $>$75\% of the HX background (e.g.\ Mushotzky \etal 2000,
Giacconi \etal 2000 and Hornschemeier \etal 2001). Comparisons to deep sub-mm surveys have
shown that these HX sources are almost always unassociated with sub-mm
sources (e.g. Hornschemeier \etal 2000; Severgnini \etal 2000;
Fabian \etal 2000, Barger \etal 2000), which instead appear to be star forming
systems. Due to the expected association between HX and MIR emission a
closer correlation should be found in the mid-IR band.

We present here {\it BeppoSAX} hard X-ray observations of the S1
region of the European Large Area {\it ISO} Survey (ELAIS) at
15$\mu$m. ELAIS (Oliver \etal 2000) was the largest program
undertaken by {\it ISO} and, covering an area of $\sim$12 sq. deg, is
the largest 15$\mu$m (Serjeant \etal 2000) and 90$\mu$m (Efstathiou
\etal 2000) survey before {\it SIRTF} and {\it FIRST} come into
play. In this paper we focus on observations in the Southern S1
region. In addition to the {\it ISO} observations this region has been
covered in the U, B, R and I (La Franca \etal in prep., Heraudeau
\etal in prep.) optical bands and the 1.4 GHz radio (Gruppioni \etal
1999) band. Selected objects have been observed spectroscopically in
the optical (La Franca \etal in prep., Gruppioni \etal in prep.) and
photometrically in the near-IR (Heraudeau \etal in prep.). The field
of view of the {\it BeppoSAX} MECS (25 arcminutes of radius; Boella et
al, 1997a,b) matches well with that of the ELAIS sub-fields, covering
a considerably larger area than both {\it Chandra} and {\it
XMM-Newton} (see Fig. 1). It is thus ideal for a shallow X-ray
survey of ELAIS fields.  Throughout this paper $H_0$=50 kms$^{-1}$,
$\Lambda$=0 and $q_0$=0.5 are used.

%
\section{OBSERVATIONS} 
%

The {\it BeppoSAX} observations were taken with both the LECS and MECS
instruments. The LECS observations are not considered here due to
their lower sensitivity and much poorer PSF below 2 keV.

The MECS observations (Boella \etal 1997b), from 1.3 to 10 keV, cover
5 of the 9 sub-fields in the S1 region of ELAIS (see Fig. 1). The
total area covered is $\sim$1.7 sq.deg., although the sensitivity of
MECS varies with off-axis distance with the most sensitive
observations covering 0.7 sq.deg. The on-source exposure time for
these 5 pointings was on average $\sim$36 ksec, corresponding to a
flux limit of $\sim$10$^{-13}$ ${\rm erg}\, {\rm s}^{-1}\, {\rm
cm}^{-2}$ on-axis (see Table 1).

The {\it BeppoSAX} positional accuracy is dependent on many factors
(see Fiore \etal 2000b for details) leading to a 95\% error radius of
one arcmin at off-axis distances $<$12 arcmin and 1.5 arcmin at
greater distances. Slightly better accuracies have been achieved in those
fields where we have been able to identify more than one source at
another wavelength and perform a positional shift of the X-ray sources to match the more accurate optical positions (see Table 1).

%
\section{DATA}
%

\subsection {ELAIS data}

As discussed in the introduction, the characteristics of the ELAIS
selection and the IR counts have been presented by Oliver et
al. (2000) and Serjeant \etal. (2000). The survey has covered an area
of $\sim$ 12 deg$^2$ down to $\sim$1 mJy in the 15$\mu$m lw3 filter (referred to here as MIR, Cersarksky et al, 1996). In this paper
we focus on observations in the S1 region centered at $\alpha$(2000) =
00$^h$ 34$^m$ 44.4$^s$, $\delta$(2000) = -43$^{\circ}$ 28$^{\prime}$
12$^{\prime \prime}$, which covers an area of about $2^{\circ}
\times 2^{\circ}$. About 100 sources of the Preliminary Analysis of ELAIS (Serjeant \etal 2000)
down to R$\sim$20.5 mags have been spectroscopically identified during three spectroscopic
identification campaigns in 1998 (La Franca \etal in prep., Gruppioni
\etal in prep).

\subsection{{\it BeppoSAX} data}

The {\it BeppoSAX} observations were cleaned and linearized using the
SDC on-line analysis. The MECS2 and MECS3 images were co-added
together and binned by a factor of 2.  Source detection was performed
using XIMAGE with the interactive routine {\it sosta}. When measuring
the counts of a source a box of 13 pixels (each pixel corresponds
to 8 arcsecs) was used. This was found to maximize the signal
to noise. As the PSF is much wider than the positional uncertainty, the effect on the flux from poor centroiding with weak sources is negligible. The net counts were automatically corrected for vignetting, sampling dead time
and the point spread function. For each source the background was
measured in large apertures at nearby orthogonal sky positions.

The conversion from counts to fluxes was determined using a power law
of $\alpha$=0.7 ($f_{\nu}\propto\nu^{-\alpha}$). We experimented with
a range of column densities (20$<$log(N$_H$)$<$23) but found small
differences ($\sim$10\%) in the conversion factors as compared to
the uncertainty in the counts of the X-ray sources and the flux
uncertainties of the MIR sources, which are of the order of 30\%
(Serjeant \etal 2000). The overall conclusions are not affected by
these uncertainties.

We initially found 13 sources at the SNR$>$3 Poisson level in either
the 1.3 to 10 keV (T) or 2 to 10 keV (HX) band although 2 of these sources
could be unreliable because of possible source confusion (sources s2 and s4), and are included in
a supplementary table (see Tables 2 and 3 and Fig. 1). Based on the
source box size we would expect $<$1 spurious source due to background
fluctuations. 
Considering the relatively small cosmic volume sampled, the SNR$_{HX}$$>$3 number counts of sources are reasonably consistent with
those found by Giommi, Perri and Fiore (2000), see Fig. 2. We have
added to the supplementary list, Table 3, 2 sources just below the detection
threshold, with SNR$_{T}$=2.9, but which are
close ($\sim$60 arcsec) to the position of an ELAIS 15$\mu$m QSO. These
sources are used in all the further analysis.

The softness ratios of the sources were determined following the
procedure of Fiore \etal (2000b) with the ratio defined as (S-H)/(S+H)
where S=count rate in 1.3 to 4.5 keV band and H=count rate in 4.5 to 10 keV band, see Fig. 3 and
Tables 2 and 3. The distribution of ratios is consistent with that
found in the {\it BeppoSAX} HELLAS survey (i.e. $-$1$<$(S-H)/(S+H)$<$1, Fiore \etal 2000b).

\subsection{Positional correlation between the {\it BeppoSAX} and
the {\it ISO} sources}

We have positionally correlated the HX sources to the MIR sources
finding 6 matches (Table 4, Fig. 4). One further match is made with a 2.6$\sigma$ {\it ISOCAM} source. This source is included in the further analysis although it should be noted that this is not a significant {\it ISOCAM} detection. The separation between the X-ray and optical centroids are $<$ 60.5 arcsecs in all cases. The optical counterparts of the {\it ISOCAM} sources have magnitudes of R$<$20 mags. This optical magnitude limit is consistent with that found for $f_{HX}>$10$^{-13}$ ${\rm erg}\, {\rm s}^{-1}\, {\rm cm}^{-2}$, the depth of this survey, in other HX surveys (e.g.\ Schmidt et al, 1998, Akiyama et al, 2000 and Hornschemeier et al, 2001). 

The surface density of the {\it ISOCAM} sources is $\sim$150 sources/deg$^2$, and
therefore the chance of a mis-association with a {\it BeppoSAX} error
box (60 arcsec radius) is 13\%. In previous {\it BeppoSAX} and {\it ASCA} surveys (e.g.\ Akiyama \etal, 2000 and La Franca \etal 2000)
sources have been matched to optical counterparts. In these studies the high optical source density often results in $>$1 possible optical counterpart within the large {\it BeppoSAX} and {\it ASCA} error boxes. The relatively low source density of MIR in this study results in less ambigious source matching although we are implictly assuming that an {\it ISOCAM} source within a {\it BeppoSAX} error box is associated with the HX source. This association is expected within the Unified model of AGN (Antonucci, 1993) and is found from both observational and theoretical tests of this model (e.g.\ B95, Granato, Danese and Franceschini, 1997, Krabbe, Boker and Maiolino, 2000 and Alonso-Herrero \etal, 2001). These X-ray and infrared observations are two orders of magnitude deeper than those of B95 (see Fig. 5).

\subsection{Hard X-ray upper limits of {\it ISO} sources}

We measured 3$\sigma$ X-ray upper limits for all the spectroscopically
classified MIR sources, labeled in Fig 1. This was performed by measuring the
counts at an MIR source position in the same fashion as for the
detected sources. The measured upper limits depend on the off-axis
distance and range from 10$^{-13}$ ${\rm erg}\, {\rm s}^{-1}\, {\rm
cm}^{-2}$ on-axis to 3-5x10$^{-13}$ ${\rm erg}\, {\rm s}^{-1}\, {\rm
cm}^{-2}$ off-axis.

%
\section{DISCUSSION}
%

In this discussion we comment on the HX-MIR source identifications and
compare the observed HX-MIR properties of the {\it BeppoSAX} and {\it ISOCAM}
objects to those of other samples previously reported in the
literature (e.g. B95 and Elvis \etal 1994, hereafter E94).

\subsection{The spectroscopic object identifications}

As already discussed we found 7 MIR sources (including the 2.6$\sigma$ {\it ISOCAM} source) within the
16 error-boxes ($\sim$60 arcsec, 95\% significance, see section 2) of the {\it BeppoSAX} detections.  We have obtained optical
spectroscopy for 6 of these sources (La Franca \etal in
prep. and Gruppioni \etal in prep.), see Table 4. As the optical spectroscopic campaign targetted those sources with 17$<R<$20.5 mags, we did not obtain a spectroscopic identification for the brightest source. We find 4 normal broad emission line QSOs, with 0.4$<z<$2.6 and $10^{44}<L_{HX}<10^{46}$ erg
s$^{-1}$ and 1 Broad Absorption Line QSO (BALQSO) at $z$=2.2 and $L_{HX}$=6x10$^{45}$ erg s$^{-1}$. The 2.6$\sigma$ {\it ISOCAM} source is associated with a
galaxy at $z$=0.325 and has $L_{HX}$$\sim 10^{44}$ erg
s$^{-1}$ if the association is real. Based on the distribution of object types found in other {\it BeppoSAX} and {\it ASCA} surveys (e.g.\ Akiyama et
al, 2000 and La Franca \etal 2000) we would expect to find two Type 2 sources in our study. We find none which seems surprising but is not significant (1$\sigma$) given the small number of sources. One
of the two low redshift QSOs ($z<$1) is radio detected (source 3,
$z$=0.559, $f_{1.4GHz}$ = 1.5 mJy, Gruppioni \etal 1999). BALQSOs are
rarely detected in the X-ray, particularly in the soft band,
possibly due to large absorption of the
X-ray emission (Gallagher \etal 1999). This is the first time a high
redshift BALQSO has been detected in both the MIR and HX.

The unclassified source 8 is a bright optical source with an extended optical profile and is possibly detected at 90$\mu$m ($f_{90\mu m}$ = 81 mJy,
Efsthathiou \etal 2000). Its soft HX ratio (section 4.2) is consistent with either an AGN or a thermal Bremsstrahlung emitting source. Its high HX/MIR flux ratio (section 4.3) suggests the former and it is probably a nearby Seyfert galaxy.

The detection of a normal galaxy seems surprising although many apparently normal galaxies have been detected in recent {\it Chandra} surveys (e.g. Mushotzky \etal 2000, Fiore \etal 2000a, Brandt \etal 2000, Giacconi \etal 2000 and Hornschemeier \etal 2000) although none have such a high X-ray luminosity and many tend to have harder X-ray spectral slopes. This object has an HX/MIR flux ratio, X-ray spectral slope and luminosity similar to QSOs although lacks the typical QSO broad optical emission lines. Other interesting possibilities are that it is a BL Lac object, although the lack of radio emission and a strong Calcium break suggest against this, or a dominant cD cluster galaxy with the X-ray emission coming from the cluster medium. The expected log(HX/MIR) ratio for this latter possibility is $-$4.7$\pm$0.4 (e.g. Edge and Stewart, 1991 and Bregman, McNamara and O'Connell, 1990). Due to the uncertain nature of this source these possibilities are not further pursued.

\subsection{Softness ratios}

We have compared the softness ratios (see Fiore \etal 2000b) to those expected for various HX SEDs, see Fig. 6. The softness ratios of the QSOs, although
extremely uncertain, show a variety of soft and hard sources 0.5$<$(S-H)/(S+H)$<$-0.5, similar to those found in the {\it BeppoSAX} HELLAS survey (Fiore \etal 2000b).

\subsection{The HX/MIR flux ratios}

We have compared the observed HX/MIR flux ratios to
those expected for a variety of object types. B95 statistically studied
the correlation between HX and 12$\mu$m for the {\it IRAS} 12$\mu$m
sample of AGN and emission line galaxies (Rush, Malkan, Spinoglio, 1993, hereafter RMS). The 12$\mu$m sample represents the properties of MIR sources in
the local universe whilst the redshifts of these
{\it ISOCAM}/{\it BeppoSAX} objects are substantially higher. Therefore in order to compare this study to that of B95 it is necessary to take into
account the effects of redshift and the construction of spectral
energy distributions (SEDs) are necessary. In the generation of SEDs we have decided to take an empirical approach due to the uncertain contribution of AGN and galactic activity at mid-IR wavelengths (e.g.\ see the detailed modelling of Cen A which is clearly an AGN although its infrared emission appears to be dominated by galactic processes, Alexander \etal, 1999). To do this we firstly evaluated and constructed HX and infrared SEDs of QSOs, Seyfert 2 and HII galaxies and then normalized the X-ray SED to reproduce the HX/MIR flux ratios of the
local value at $z\sim$0 as determined by E94 for QSOs, and by B95 for Seyfert
2 and HII galaxies. Only a few Seyfert 2s were detected at HX energies by B95 although the HX/MIR flux ratio distribution is very similar to the {\it IRAS} 60$\mu$m selected, HX detected 16 Seyfert 2 galaxies in Alexander (2001) (log(HX/MIR)=-7.0$\pm$0.7). The B95 HX/MIR flux ratio for the HII galaxies is an upper limit.


The HX SEDs were taken from Pompilio, La Franca and Matt (2000) for
the cases of QSO and Seyfert 2; the QSO SED is a two part power law whilst the Seyfert 2 SED uses this same spectrum convolved with the Seyfert 2 absorbing column density
distribution found by Maiolino \etal (1998). The Seyfert 2 galaxies in
the Maiolino \etal study were not MIR selected. However, the column
density distribution is very similar to that found for far-IR
selected Seyfert 2 galaxies (i.e.\ log($N_H$)$\sim$22 to 25 cm$^{-2}$, Alexander, 2001). The HII galaxy HX SED
was determined assuming a Bremsstrahlung thermal spectrum with kT=5.8
keV, as found for the starburst galaxy NGC253 (Cappi \etal 1999).


The Seyfert 2 and HII galaxy infrared SEDs were determined using the
Xu \etal (1998) empirical algorithm which takes the four {\it IRAS} band
fluxes to predict an overall 2 to 120$\mu$m SED assuming three basic
components: AGN, starburst and cirrus. The $<$12$\mu$m emission is predicted from the {\it IRAS} colours and produced using AGN, starburst and cirrus observational templates which include PAH emission and dust absorption features. The sample used to determine
these SEDs was the RMS sample as classified by Alexander and Aussel
(2000). These SEDs do not account for starlight, which can potentially
be a large contributor for $\lambda$$<$7$\mu$m and would lead to underestimates of the
HX/MIR ratios. Therefore we have only used these SEDs for z$<$1,
which is sufficient for the non-QSO sources. For the QSOs we have used
the empirically determined SED of E94
which includes the contribution from starlight and gives a good match
to the low redshift optically selected PG QSOs (Schmidt and Green,
1983). As this SED is derived mostly from low redshift sources, we are assuming little SED evolution for z$<$3 (see section 4.3.1 for discussion).

The HX/MIR flux ratios of the {\it BeppoSAX}/{\it ISO} sample are shown
in Fig. 7a. The errorbars associated with the SEDs take into account
the statistical spread in the HX/MIR ratios measured locally by B95. In the case of Seyfert 2s and QSOs the effect of redshift is to
increase the HX flux with respect to the MIR flux. The difference with
redshift is not dramatic for either object type because the assumed HX
column densities are quite low, although see section 4.3.2 for the case
of a Compton thick source (log(N$_{H}$)$>$24 cm$^{-2}$). The HII galaxies are difficult to detect in
the HX at any redshift but particularly at high redshift where the HX
K-correction is positive. In the case of a higher temperature of the X-ray emitting gas than that assumed here,
the HX/MIR ratio will stay constant to higher redshifts, dropping off
significantly at the exponential cut-off energy (i.e. at
$z=(T_{HX}/2)-1$, where $T_{HX}$ is the Bremsstrahlung temperature in
keV).

The flux ratio distribution of the HX/MIR upper limits are shown in Fig. 7b. The mean HX upper limit and log(HX/MIR) are
3.3x10$^{-13}$ ${\rm erg}\, {\rm s}^{-1}\, {\rm cm}^{-2}$ and $-$5.1 respectively.
To detect all the sources requires significantly deeper
X-ray observations. HX observations of the Northern
ELAIS fields will be carried out by {\it Chandra} and {\it XMM-Newton}
to limiting fluxes of $\sim$3x10$^{-15}$ and 10$^{-14}$ ${\rm erg}\, {\rm s}^{-1}\, {\rm cm}^{-2}$
respectively. Assuming the mean flux and HX/MIR flux ratio of this {\it BeppoSAX} survey, the {\it Chandra} and {\it XMM-Newton} surveys should reach mean flux ratios of log(HX/MIR)$\sim$$-$7.1 and log(HX/MIR)$\sim$$-$6.6 although clearly some objects with lower ratios will also be detected. The {\it Chandra} survey (two ACIS-I 16 arcmin pointings) covers a small area but should detect all the AGN and some HII galaxies within the field of view whilst the {\it XMM-Newton} survey covers a similar area to this survey and should detect all the QSOs, most of the Seyfert 2 galaxies and a few HII galaxies.

\subsubsection{The QSOs}

The majority of the HX-MIR associated sources are QSOs. The observed
QSO HX/MIR flux ratios appear consistent with that of the median E94 QSO SED.
The E94 UVSX sample includes 47 quasars selected to have at least 300
counts in the Einstein IPC and with V$<$17 mags to be observable by IUE; 29
members are radio quiet and 18 radio loud.  The sample is consequently
biased toward objects at low redshifts (although 7 objects have $z>$1,
4 $z>$2, 1 $z>$3), of moderate luminosity and with high X-ray to optical flux
ratios.

In order to increase the statistics of our comparison we added to the MIR sample QSOs from Andreani, Franceschini and Granato (1999) with $f_{12\mu
m}>$200 mJy, and created as a comparison an optically selected sample using 12$\mu$m detections or upper limits of the PG
optically selected QSOs from Sanders \etal (1989). To increase the
number of objects at high redshift we have taken those objects from
Neugebauer \etal (1986) with z$>$2. We have determined which of these
are BALQSOs using the list of Junkkarinen, Hewitt and Burbidge
(1991). The data are shown in Fig. 8. The mean log(HX/MIR) ratios of
the low redshift QSOs (z$<$0.6) are $-$5.7$\pm$0.7 and $-$5.6$\pm$0.3 for
the MIR and optically selected objects respectively. These flux ratios
are consistent with each other and with the E94 SED. This suggests
that MIR selected QSOs of R$<$20 mags come from the same population as optically selected QSOs and that there is little SED evolution for z$<$3.

As a further and more definitive test we have compared the log(B/MIR)
ratios of all the MIR detected QSOs in the S1 region to
those of the PG QSO sample (Schmidt and Green, 1983 and Sanders et al, 1989), see Fig. 9. The advantage of this
comparison is that we can use the full QSO set from both the ELAIS and
PG surveys. The MIR QSOs are consistent with the flux ratios of the
E94 SED and the PG QSOs. This further strengthens the evidence that MIR
selected QSOs of R$<$20 mags come from the same population as optically selected
QSOs.

\subsubsection{The BALQSO}

Although the statistics are still poor, low redshift BALQSOs seem to be located in
a different region of the HX/MIR plane to QSOs, see
Fig. 8. This is consistent with that expected from an
absorbed source. The low-redshift QSO, that lies close to the BALQSOs
in the HX/MIR plane shown in Fig. 8, is a narrow line Seyfert 1
galaxy (IZw1), an unusually red source (e.g.\ see E94) which may be different to the general QSO population
(e.g. see Brandt and Gallagher, 2000). Our BALQSO has a HX/MIR flux
ratio similar to that found for normal QSOs, apparently in
contradiction with that found for low redshift BALQSOs. This
difference could be due to the negative HX K-correction effect of an
absorbed HX spectrum, see Fig. 6. We can test whether the observed HX/MIR flux ratio is
compatible with that expected for an absorbed HX source by
constructing a possible BALQSO SED. We have chosen to use the Compton
thick HX spectrum of the Type 2 QSO object IRAS09104+4109
(Franceschini \etal 2000) and the QSO infrared SED of E94, normalised to the HX/MIR ratio of IRAS09104+4109. The positive effect of redshift on the HX emission can
clearly be seen (see Fig. 8) showing that a high redshift HX
absorbed source can have a similar HX/MIR flux ratio to an unabsorbed
source. Therefore, in contrast to low redshift BALQSOs, high redshift
BALQSOs should be comparatively easy to detect in the HX band,
allowing a determination of their true fraction in the high-z QSO population.

%
\section{CONCLUSIONS}
%

We have presented shallow {\it BeppoSAX} observations, reaching an on-axis
($\sim$0.7 sq. deg) 2-10 keV sensitivity of $\sim$10$^{-13}$ ${\rm erg}\,
{\rm s}^{-1}\, {\rm cm}^{-2}$, of the Southern S1 region of ELAIS,
reaching a 15$\mu$m sensitivity of 1 mJy. This is the first HX analysis
of an {\it ISOCAM} survey. We
have constructed HX and infrared SEDs to determine the expected flux
ratios for QSOs and BALQSOs up to $z$=3.2 and for Seyfert 2s and normal
galaxies up to $z$=1.0. Our main findings are: 

(i) we detect 13 sources with SNR$>$3 in the 1.3-10 keV or 2-10 keV
X-ray bands and a further 2 sources with less reliable detections that have positions close ($\sim$60 arcsec) to QSOs. The number
densities of the SNR$_{HX}>$3 sources are consistent with the {\it ASCA}
and {\it BeppoSAX} logN-logS function.

(ii) 6 of these sources have a reliable {\it ISOCAM} counterpart and one further source has a less reliable (2.6$\sigma$) {\it ISOCAM} counterpart.  
We have optical spectroscopic classifications for 6 of these sources
finding 4 QSOs, 1 BALQSO at $z$=2.2 and 1 apparently normal galaxy (the 2.6$\sigma$ {\it ISOCAM} source). The unclassified object has X-ray and photometric properties consistent with that of a nearby Seyfert galaxy. The galaxy has properties suggesting either an unusual QSO or a galaxy cluster.

(iii) the QSOs cover a wide redshift range (0.4$<$z$<$2.6), and have HX/MIR flux ratios
consistent with those found for nearby {\it IRAS} and optically selected PG
QSOs and the QSO SED of Elvis et al (1994). By further comparing the B/MIR flux ratios of the MIR
QSOs to those of the blue band selected PG sample, we suggest that MIR selected QSOs of R$<$20 mags come
from the same population as optically selected QSOs.

(iv) the high redshift BALQSO has a HX/MIR ratio similar to that of
QSOs, but different to that found for low redshift BALQSOs. This
difference can be explained as the negative K-correction effect of an
absorbed X-ray spectrum observed at high redshift. This suggests that
high redshift BALQSOs should be comparatively easy to detect in the HX
band, allowing the true fraction of BALQSOs in the high redshift QSO population to be
determined.

%
\acknowledgements
%

We acknowledge the EC TMR network (FMRX-CT960068) for financial support and MURST Cofin-98-032 and ASI contracts for partial support. DMA additionally thanks the NSF CAREERS grant AST-998 3783 for post-doctoral support. This research has made use of SAXDAS (SAX Data Analysis System) linearized and cleaned event files
(Rev.2.0) produced at the {\it BeppoSAX} Science Data Center and the
NASA/IPAC Extragalactic Database (NED) which is operated by the Jet
Propulsion Laboratory, California Institute of Technology, under contract
with NASA. We gratefully thank Belinda Wilkes, the referee, for an efficient and thorough reading of this manuscript. We further thank Cong Xu for providing the code to determine the
mid-IR SEDs used in this paper and Sarah Gallagher, Ann Hornschemeier and Niel Brandt for valuable comments on earlier drafts of this paper.


%

%

%
%

\clearpage

%
%

\begin{table}
\caption{{\it BeppoSAX} MECS observations.}
\label{TABLE:obs}
\footnotesize
\setlength{\tabcolsep}{0.6mm}
\begin{tabular}{lccccc}
\hline
\multicolumn{1}{c}{Field}&{Centre position}&{Exp}&{Date}&{X-offset}&{Y-offset}\\
\multicolumn{1}{c}{(1)}&{(2)}&{(3)}&{(4)}&{(5)}&{(6)}\\
\hline
S1.2 & 0 30 57.3  -43 38 12.9 & 34.0 & 15/12/99 & 0.0222 & 0.0040\\
S1.4 & 0 33 43.6  -42 50 25.0 & 33.1 & 20/12/99 &   &  \\
S1.5 & 0 34 32.6  -43 29 24.0 & 36.1 & 19/12/99 &   &  \\
S1.7 & 0 37 30.3  -42 38 38.2 & 43.2 &         &         &      \\
 &            & {\it 27.8} & 18/07/99 & -0.0206 & 0.0041\\
 &            & {\it 15.4} & 19/12/99 &   &  \\
S1.8 & 0 38 26.7  -43 17 23.7 & 31.2 & 17/07/99 &  0.0000 & 0.0062\\
\hline\\
\end{tabular}

\noindent{\footnotesize {\em Notes}: Col.(1) ELAIS region; (2) uncorrected X-ray
field centre position (J2000); (3) observation exposure time (ksec); (4)
date of observation; (5) and (6) RA and Dec corrections (deg), see section 2.}

\end{table}

%
%
\begin{table*}
\tiny
\caption{{\it BeppoSAX} sources.}
\label{TABLE:detected}
\setlength{\tabcolsep}{1.5mm}
\begin{tabular}{lcccccccccccl}
\hline
\multicolumn{1}{c}{Obj. no.} & {X-ray position} & $SNR_{T}$ & ${Counts_{HX}}$ &
$SNR_{HX}$ & $f_{HX}$ & {Softness} \\
\multicolumn{1}{c}{(1)}& {(2)}& {(3)}& {(4)}& {(5)} & {(6)} & {(7)}\\
\hline
1 & 0 30 13.1  -43 53 31.3	& 5.1 &     5.7$\pm$1.2 &      4.8 & 5.3$\pm$1.2 & -0.11$\pm$0.28 \\
2 & 0 30 50.2  -43 37 15.4	& 3.6 &     1.5$\pm$0.4 &      3.7 & 1.4$\pm$0.4 & -0.51$\pm$0.39 \\
3 & 0 32 13.8  -43 32 56.9	& 3.9 &     2.2$\pm$0.7 &      3.4 & 2.0$\pm$0.7 & 0.32$\pm$0.44 \\
4 & 0 32 24.3  -42 57 15.6	& 3.1 &     2.7$\pm$1.0 &      2.8 & 2.5$\pm$1.0 & -0.39$\pm$0.56 \\
5 & 0 33 45.7  -43 18 44.7	& 3.9 &     2.7$\pm$0.7 &      3.6 & 2.5$\pm$0.7 & 0.18$\pm$0.35 \\
6 & 0 34 01.7  -42 52 38.6	& 2.9 &     1.3$\pm$0.4 &      3.3 & 1.2$\pm$0.4 & -0.37$\pm$0.47 \\
7 & 0 34 22.3  -43 18 38.3	& 2.0 &     1.5$\pm$0.5 &      3.1 & 1.4$\pm$0.5 & -0.46$\pm$0.36\\
8 & 0 35 11.7  -43 33 42.2	& 8.7 &     4.0$\pm$0.6 &      6.8 & 3.7$\pm$0.6 & 0.59$\pm$0.14\\
9 & 0 37 08.4  -42 37 07.6	& 6.5 &     2.4$\pm$0.4 &      6.0 & 2.2$\pm$0.4 & 0.31$\pm$0.20\\
10 & 0 37 14.3  -42 34 55.6	& 5.5 &     1.9$\pm$0.4 &      5.0 & 1.8$\pm$0.4 & 0.22$\pm$0.26\\
11 & 0 37 50.5  -43 27 53.1	& 3.3 &     1.8$\pm$0.6 &      2.9 & 1.7$\pm$0.6 & -0.08$\pm$0.43\\
\hline\\
\end{tabular}

\noindent{\footnotesize {\em Notes}: Col. (1) {\it BeppoSAX} source number; (2) {\it
BeppoSAX} position (J2000), see text for positional accuracy; (3) total X-ray (1.3-10 keV) SNR; (4), (5), (6)
hard X-ray (2-10 keV) count rate, SNR, flux in units of 10$^{-13}$ ${\rm erg}\,
{\rm s}^{-1}\, {\rm cm}^{-2}$; (7) softness ratio (S-H)/(S+H), see section 4.2.}

\end{table*}

%
%
\begin{table*}
\tiny
\caption{Supplementary {\it BeppoSAX} sources.}
\label{TABLE:detected}
\setlength{\tabcolsep}{1.5mm}
\begin{tabular}{lccccccl}
\hline
\multicolumn{1}{c}{Obj. no.} & {X-ray position} & $SNR_{T}$ & ${Counts_{HX}}$ &
$SNR_{HX}$ & $f_{HX}$ & {Softness} & Reason for inclusion\\
\multicolumn{1}{c}{(1)}& {(2)}& {(3)}& {(4)}& {(5)} & {(6)} & {(7)} & {(8)}\\
\hline
s1 & 0 29 57.1  -43 48 44.3	& 2.9 &     2.2$\pm$0.9 &      2.4 & 2.0$\pm$0.9 & -0.58$\pm$0.43 & SNR$_{T}$=2.9 but 33 arcsec from QSO\\
s2 & 0 32 00.4  -43 31 14.9	& 3.8 &     1.3$\pm$0.5 &      2.5 & 1.2$\pm$0.5 & 0.46$\pm$0.42 & close to source 3, maybe one source\\
s3 & 0 33 55.1  -42 55 41.8	& 2.9 &     1.0$\pm$0.4 &      2.3 & 0.9$\pm$0.4 & -0.15$\pm$0.46 & SNR$_{T}$=2.9 but 60 arcsec from QSO\\
s4 & 0 38 17.1  -43 17 22.1	& 3.4 &     1.2$\pm$0.4 &      2.4 & 1.1$\pm$0.4 & 0.32$\pm$0.44 & possibly 2 sources in HX band\\
\hline\\
\end{tabular}

\noindent{\footnotesize {\em Notes}: Col. (1) {\it BeppoSAX} source number; (2) {\it BeppoSAX} position
(J2000), see text for positional accuracy; (3) total X-ray (1.3-10 keV) SNR; (4), (5), (6) hard X-ray (2-10 keV) 
count rate, SNR, flux in units of 10$^{-13}$ ${\rm erg}\, {\rm s}^{-1}\, {\rm
cm}^{-2}$; (7) softness ratio (S-H)/(S+H), see text; (8) reason why included in the supplementary table}

\end{table*}

%
%
\begin{table*}
\tiny
\caption{{\it ISOCAM} detected {\it BeppoSAX} sources.}
\label{TABLE:detected}
\setlength{\tabcolsep}{1.5mm}
\begin{tabular}{lccccccccccccl}
\hline
\multicolumn{1}{c}{Obj. no.} & {X-ray position} & {$L_{HX}$} & HX-Optical offset&
{$f_{MIR}$}& {$L_{MIR}$} & $f_{HX}/f_{MIR}$ & Optical position & $R$ & $M_{R}$ & {z} & Classification\\
\multicolumn{1}{c}{(1)}& {(2)}& {(3)}& {(4)}& {(5)}& {(6)}& {(7)} & {(8)}& {(9)}& {(10)} & {(11)} & {(12)}\\
\hline
s1 & 0 29 57.1  -43 48 44.3 & 45.9 & 32.8 & 3.2 & 45.9 & -5.5$\pm$0.2 & 0 29 59.2  -43 48 35.3 & 17.4 &-29.4& 2.039 & QSO\\
3  & 0 32 13.8  -43 32 56.9 & 44.6 & 47.3 & 1.8 & 44.4 & -5.3$\pm$0.1 & 0 32 11.1  -43 33 22.3 & 18.7 &-24.5& 0.559 & QSO\\
s3 & 0 33 55.1  -42 55 41.8 & 45.8 & 60.2 & 0.5 & 45.3 & -5.0$\pm$0.2 & 0 33 52.8  -42 54 52.4 & 18.5 &-28.9& 2.584 & QSO\\
8  & 0 35 11.7  -43 33 42.2 &      & 60.5 & 15.8&      & -6.0$\pm$0.1 & 0 35 15.6  -43 33 57.7 & 16.4 &     &       & \\
9  & 0 37 08.4  -42 37 07.6 & 44.2 & 13.3 & 1.2$^a$ & 43.8 & -5.1$\pm$0.1 & 0 37 07.9  -42 37 18.6 & 19.4 &-22.5& 0.325 & Galaxy\\
10 & 0 37 14.3  -42 34 55.6 & 45.8 & 25.7 & 3.7 & 46.0 & -5.8$\pm$0.1 & 0 37 15.5  -42 35 14.0 & 18.3 &-28.7& 2.190 & BALQSO\\
11 & 0 37 50.5  -43 27 53.1 & 44.2 & 50.1 & 1.6 & 44.0 & -5.3$\pm$0.2 & 0 37 53.1  -43 28 24.5 & 19.2 &-23.1& 0.398 & QSO\\
\hline\\
\end{tabular}

\noindent{\footnotesize {\em Notes}: Col. (1) {\it BeppoSAX} source number; (2) {\it BeppoSAX} position (J2000), see text for positional accuracy;
(3) log
hard X-ray (2-10 keV) luminosity (${\rm erg}\, {\rm s}^{-1}$); (4) offset from the
optical source position (arcsec); (5) 15$\mu$m flux density (mJy)  from Serjeant et al
(2000), the superscript (a) refers to the 2.6$\sigma$ {\it ISOCAM} source, see section 3.3; (6) log 15$\mu$m luminosity (${\rm erg}\, {\rm s}^{-1}$); (7) HX/MIR flux ratio; (8) optical
position (J2000); (9), (10) apparent and absolute R band magnitudes (mags) from La Franca et al (in prep.); (11)
redshift; (12) source classification from La Franca et al (in prep.) and Gruppioni et al
(in prep.)}

\end{table*}


%
%

\clearpage

%
%

\figcaption{The observed ELAIS S1 sub-fields overlayed with the {\it ISOCAM}
15$\mu$m sources (points), classified sources (crosses, La Franca \etal in prep., Gruppioni \etal in prep.) and {\it BeppoSAX} detected sources (small circles). The overall field of view of the MECS instrument ($\sim$50 arcmin diameter, solid circle) and the highest sensitivity (broken circle) regions are shown.}

%

%
%

\figcaption{The cumulative 2-10 keV (HX) logN-logS
function compared to that found by Giommi, Perri and Fiore (2000). Only those sources with SNR$_{HX}>$3 are shown.}

%

%
%

\figcaption{The softness ratios of the {\it BeppoSAX}
sources as a function of HX count rate. The open circles refer to the QSOs, the filled circles refer to the BALQSO, the star refers to the galaxy and the open square refers to objects without an optical ID. The small circles refer to to the QSO supplementary sources, see Table 3 and text.}

%

%
%

\figcaption{HX to optical source matching
distances. The symbol size for both the MIR (15$\mu$m) and HX fluxes represent the source flux where a larger symbol denotes a larger flux, see Fig. 3 for object type key.}

%

%
%

\figcaption{The HX and MIR fluxes of the MIR detected
HX sources, see Fig. 3 for object type key.}

%
%

\figcaption{The softness ratios of the
MIR detected HX sources shown as a function of redshift. The dotted lines show the expected intrinsically absorbed and unabsorbed flux ratios of an AGN source and the flux ratio of a thermal Bremsstrahlung emitting galaxy or galaxy cluster. See Fig. 3 for object type key.}

%

%
%

\figcaption{Expected log(HX/MIR) 
flux ratios for QSOs, Seyfert 2s and HII galaxies over-plotted with (a) the MIR detected HX
sources and (b) the MIR sources with HX upper limits. 
The error bars are from B95 and correspond to the statistical spread in
HX/MIR colors for each source type. The crosses refer to HII galaxies and the filled squares refer to Seyfert 2 galaxies, see Fig. 3 for the object type key of the other object types. The mean depths of the {\it BeppoSAX} and
the predicted depths of the forthcoming {\it XMM-Newton} (P.I. R.  Mann)
and {\it Chandra} (P.I. O. Almaini) surveys of the Northern ELAIS regions
are shown.}

%

%
%

\figcaption{log(HX/MIR) flux ratios of MIR and optically
selected QSOs. The curves show the mean QSO and BALQSO values. The statistical spread of values are calculated from the QSOs of E94. The narrow line Seyfert 1 object IZwI is indicated as well as the two nearby BALQSOs MKN231 and IRAS07599+6508. See Fig. 7 for the object type key. The
additional HX data was taken from Turner and Pounds (1989), Della
Ceca et al (1990), Turner et al (1990), Williams et al (1992), Lawson et
al (1992), Saxton et al (1993), Sambruna et al (1994), Ceballos and
Barcons (1996), Alonso-Herrero et al (1997), Lawson and Turner (1997),
Reeves et al (1997) and Gallagher et al (1999).}

%
%

\figcaption{log(B/MIR) flux ratios of MIR and
optically selected QSOs. The curves show the mean QSO value. The statistical spread is calculated from the QSOs of E94. See Fig. 7 for the object type key. The data for the PG objects was taken from Schmidt and Green (1983) and Sanders et al (1989).}

%

\begin{figure*}
\plotone{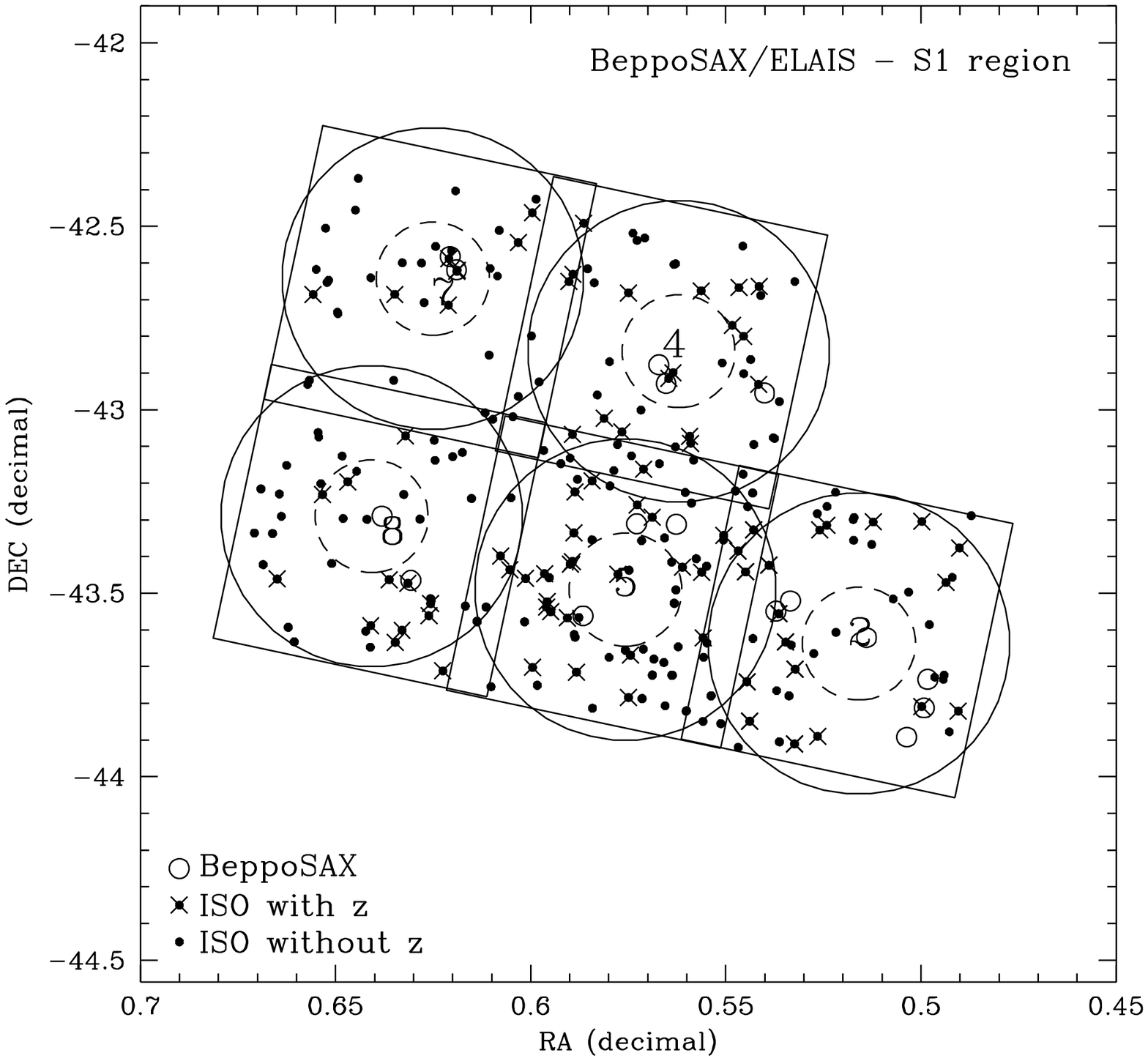}
\end{figure*}

\begin{figure}
\plotone{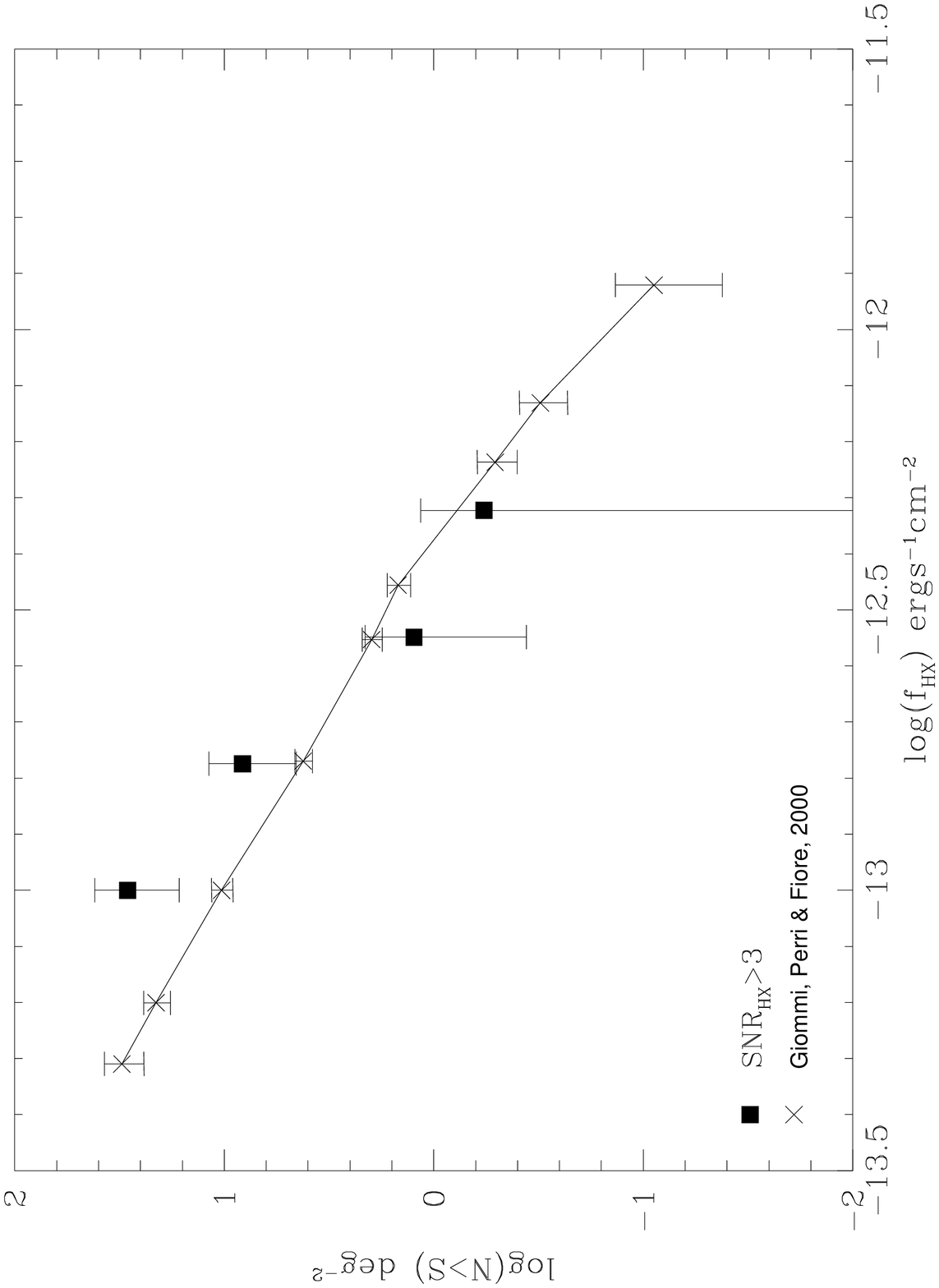}
\end{figure}

\begin{figure}
\plotone{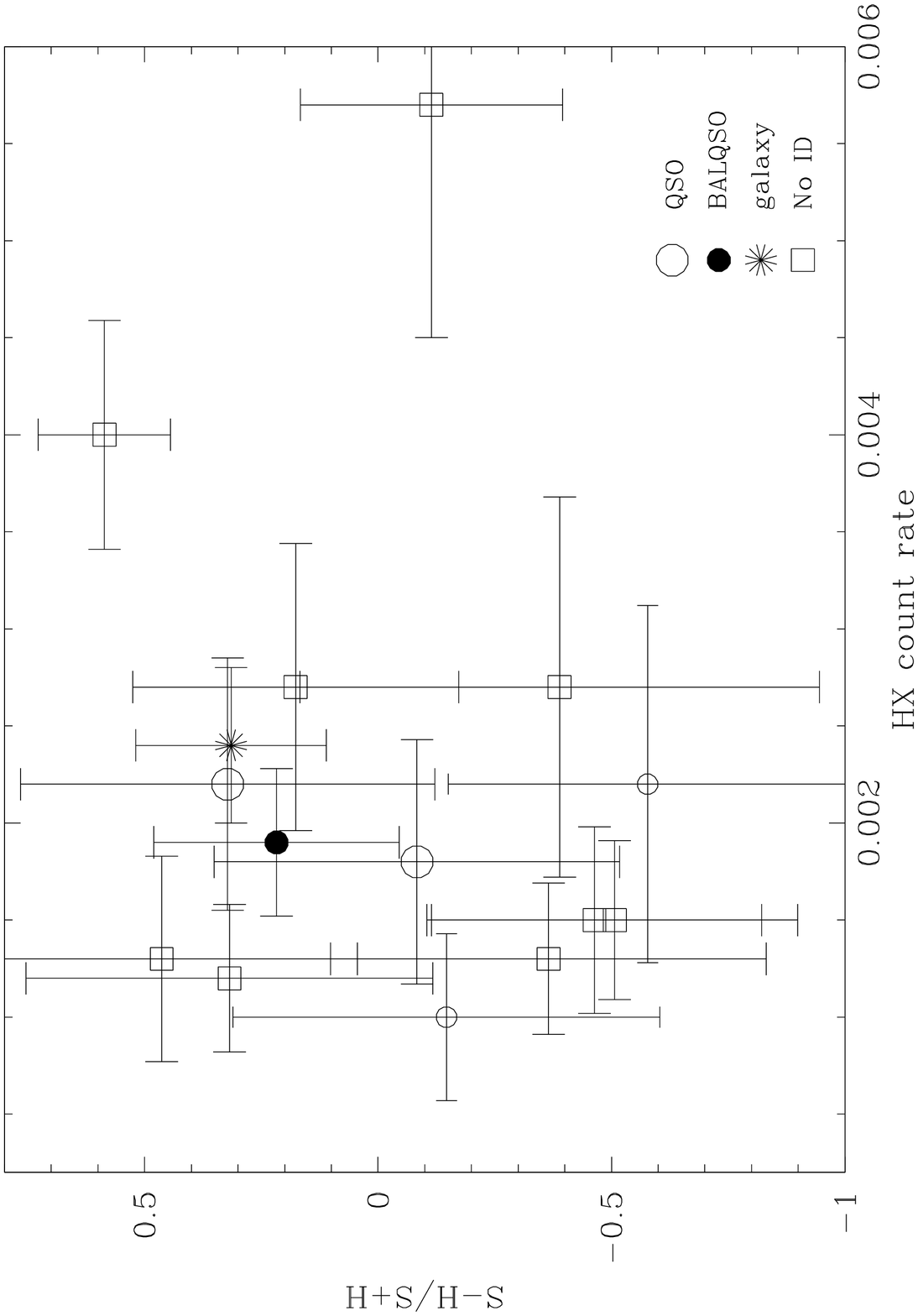}
\end{figure}

\begin{figure}
\plotone{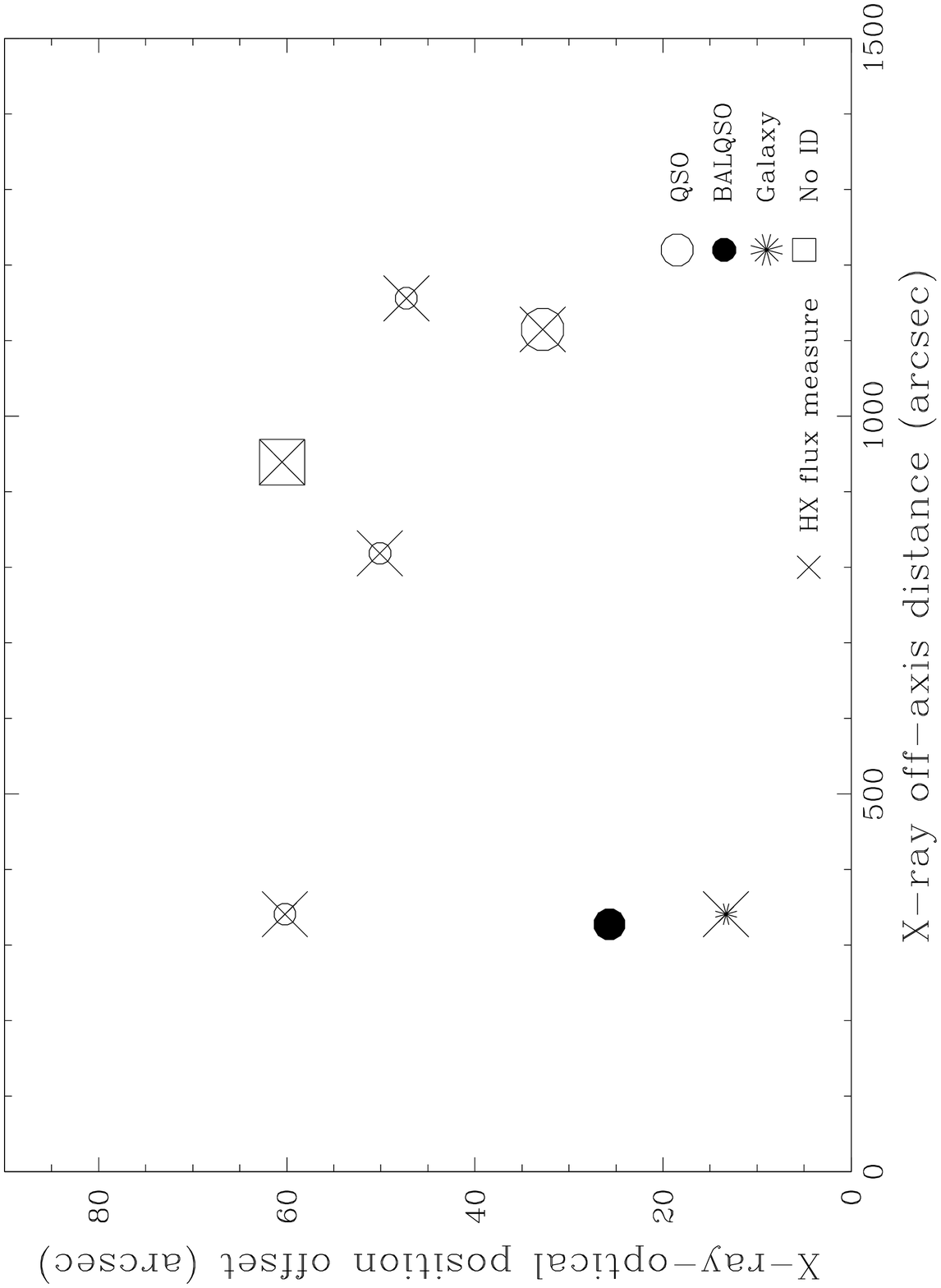}
\end{figure}

\begin{figure}
\plotone{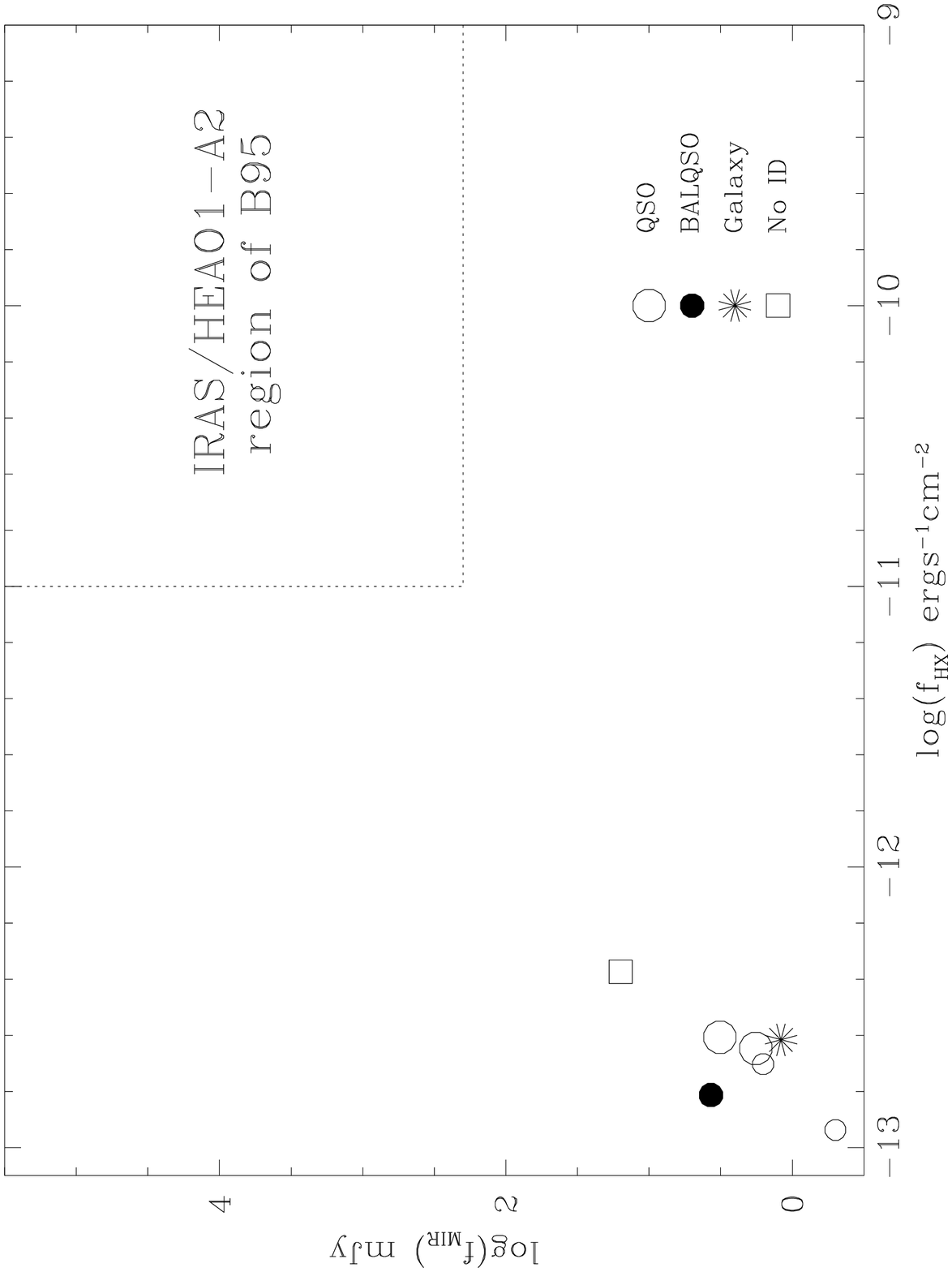}
\end{figure}

\begin{figure}
\plotone{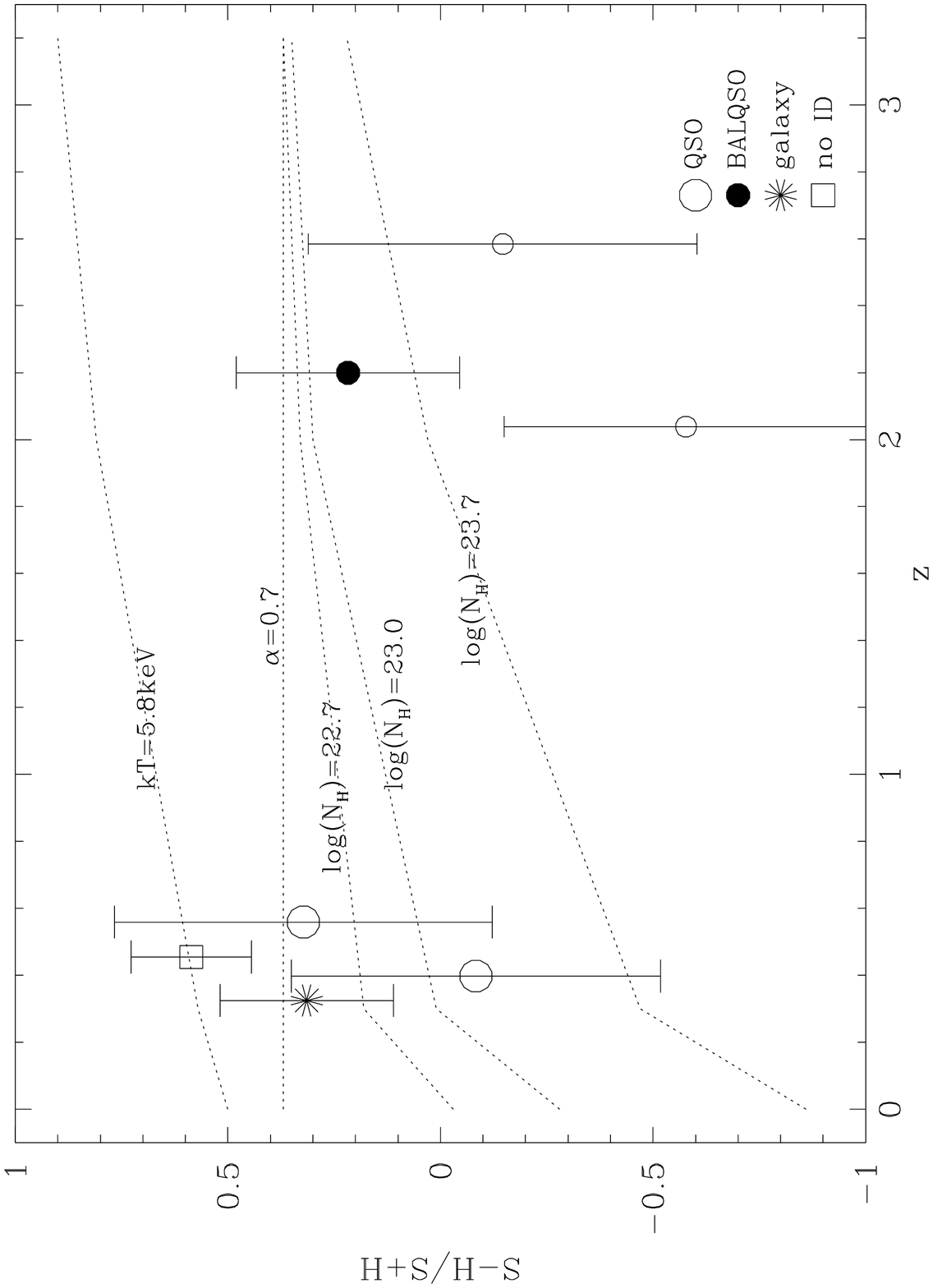}
\end{figure}

\begin{figure}
\plotone{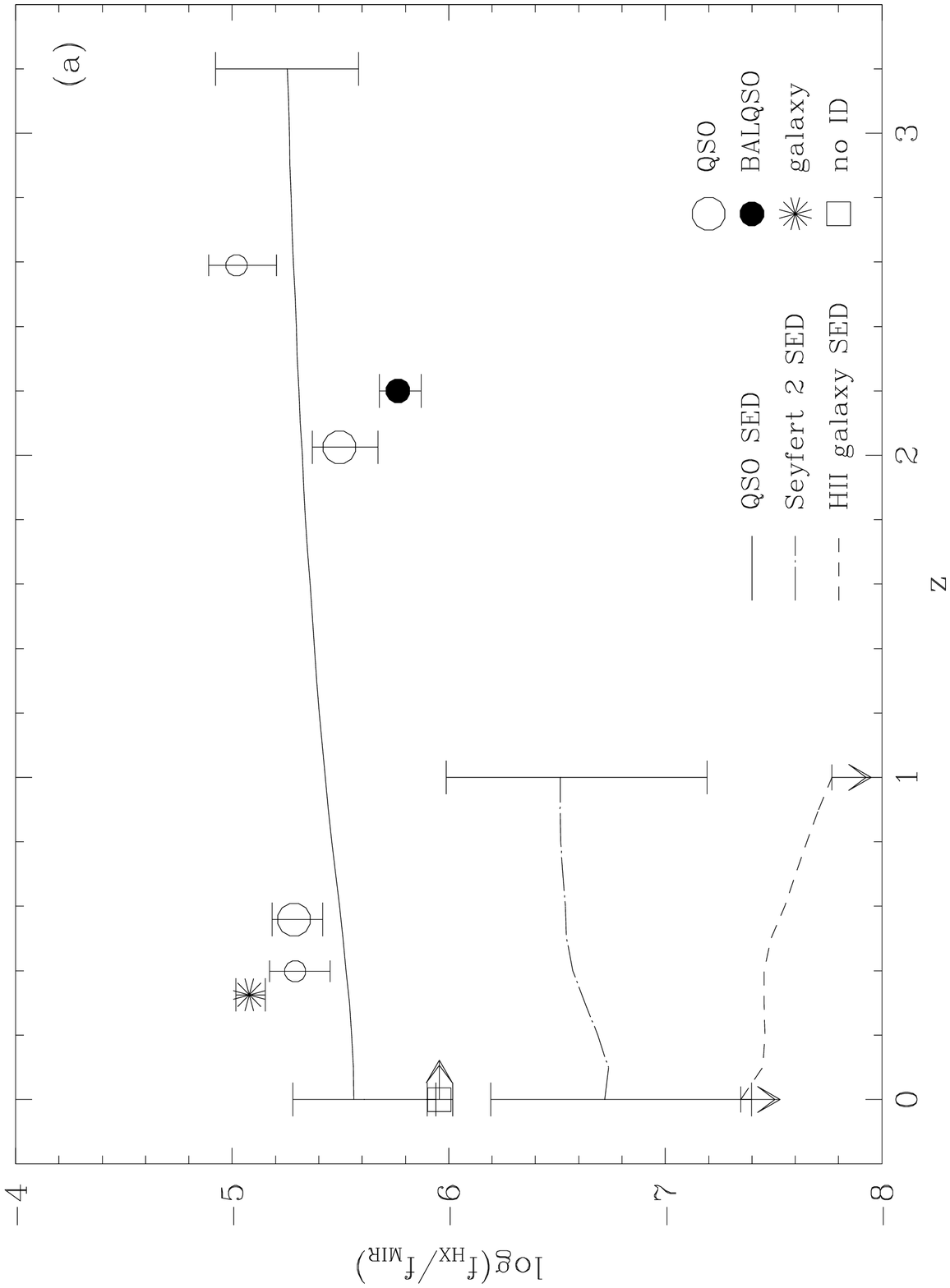}
\end{figure}

\begin{figure}
\plotone{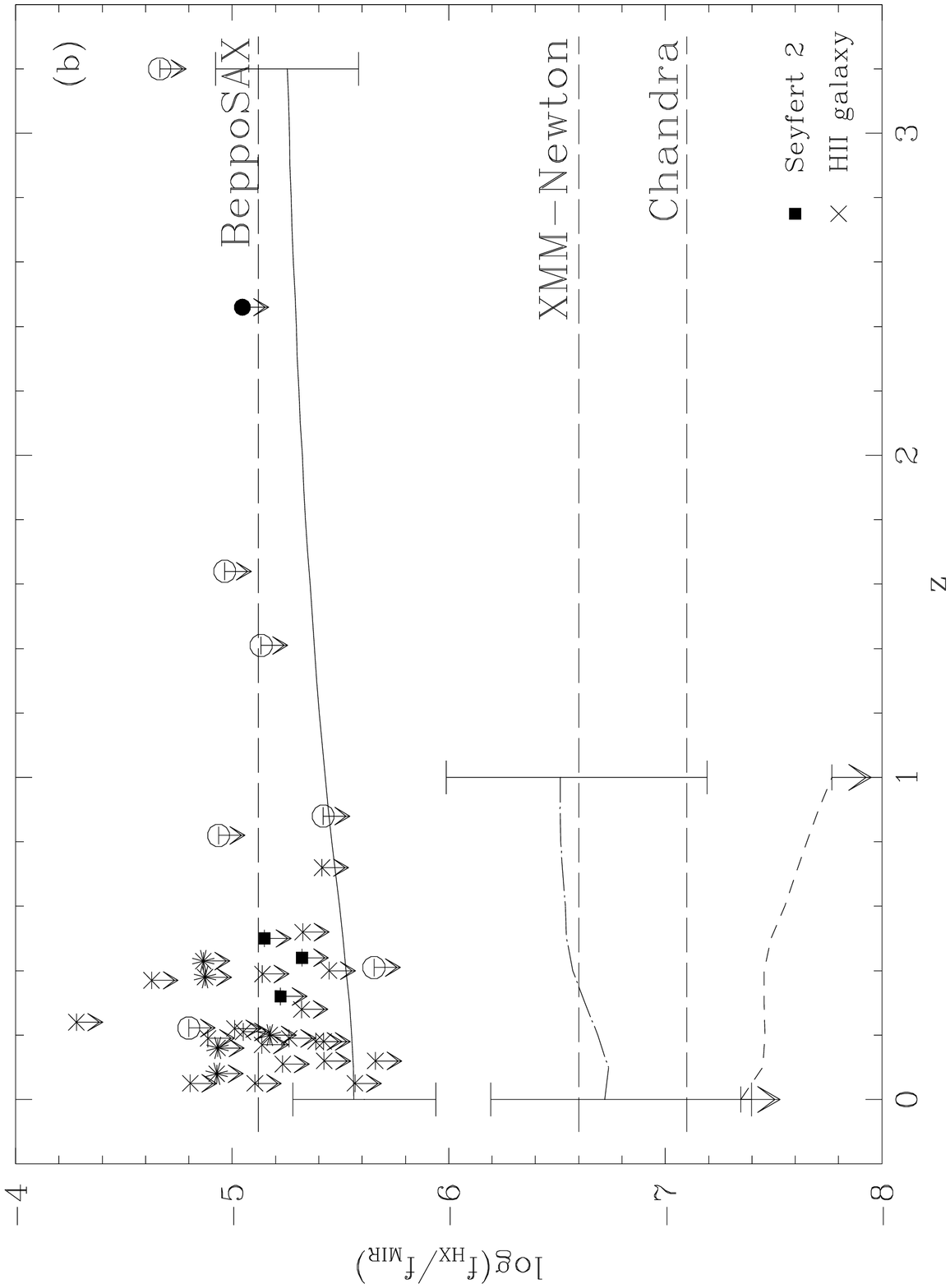}
\end{figure}

\begin{figure*}
\plotone{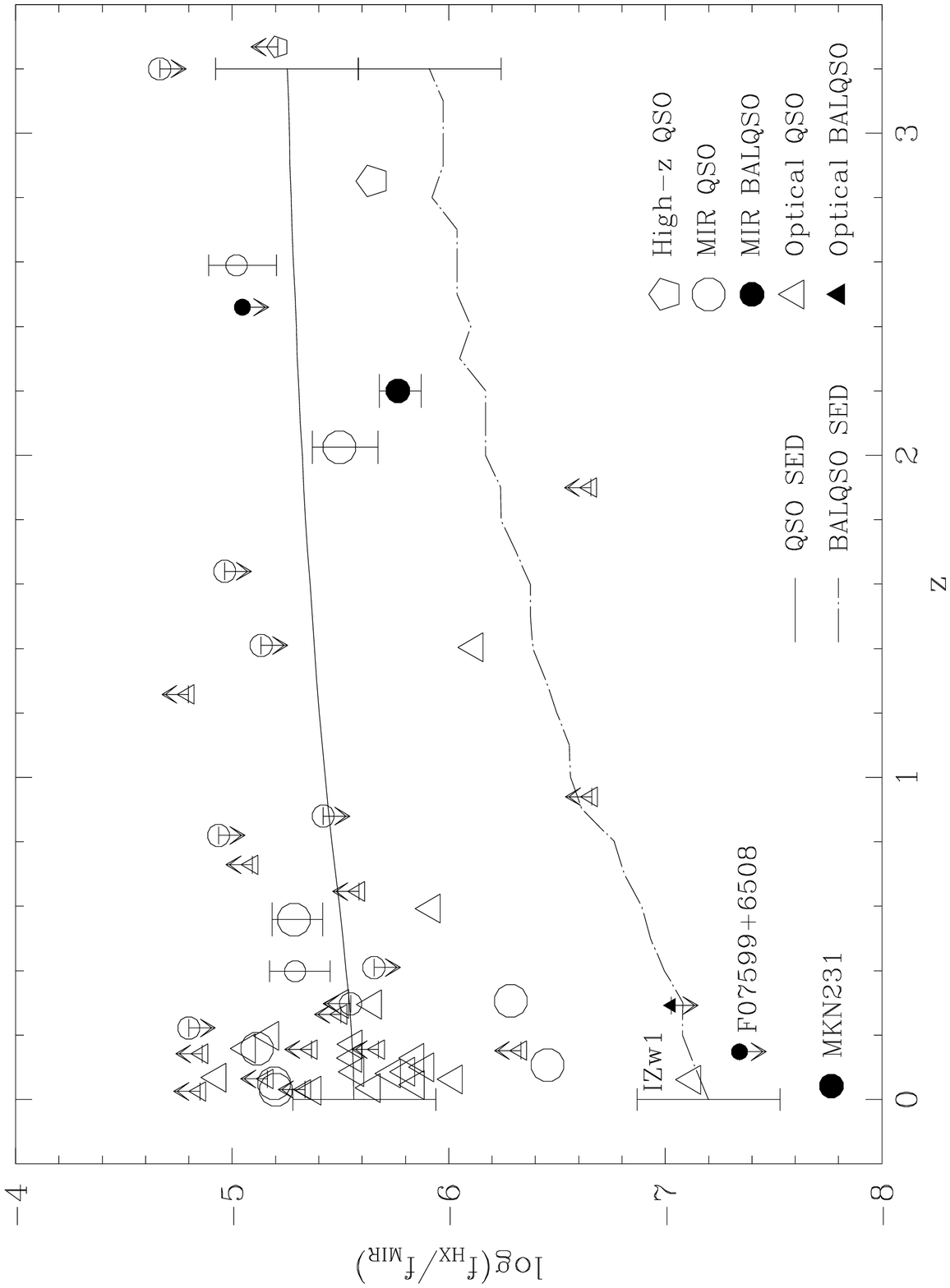}
\end{figure*}

\begin{figure*}
\plotone{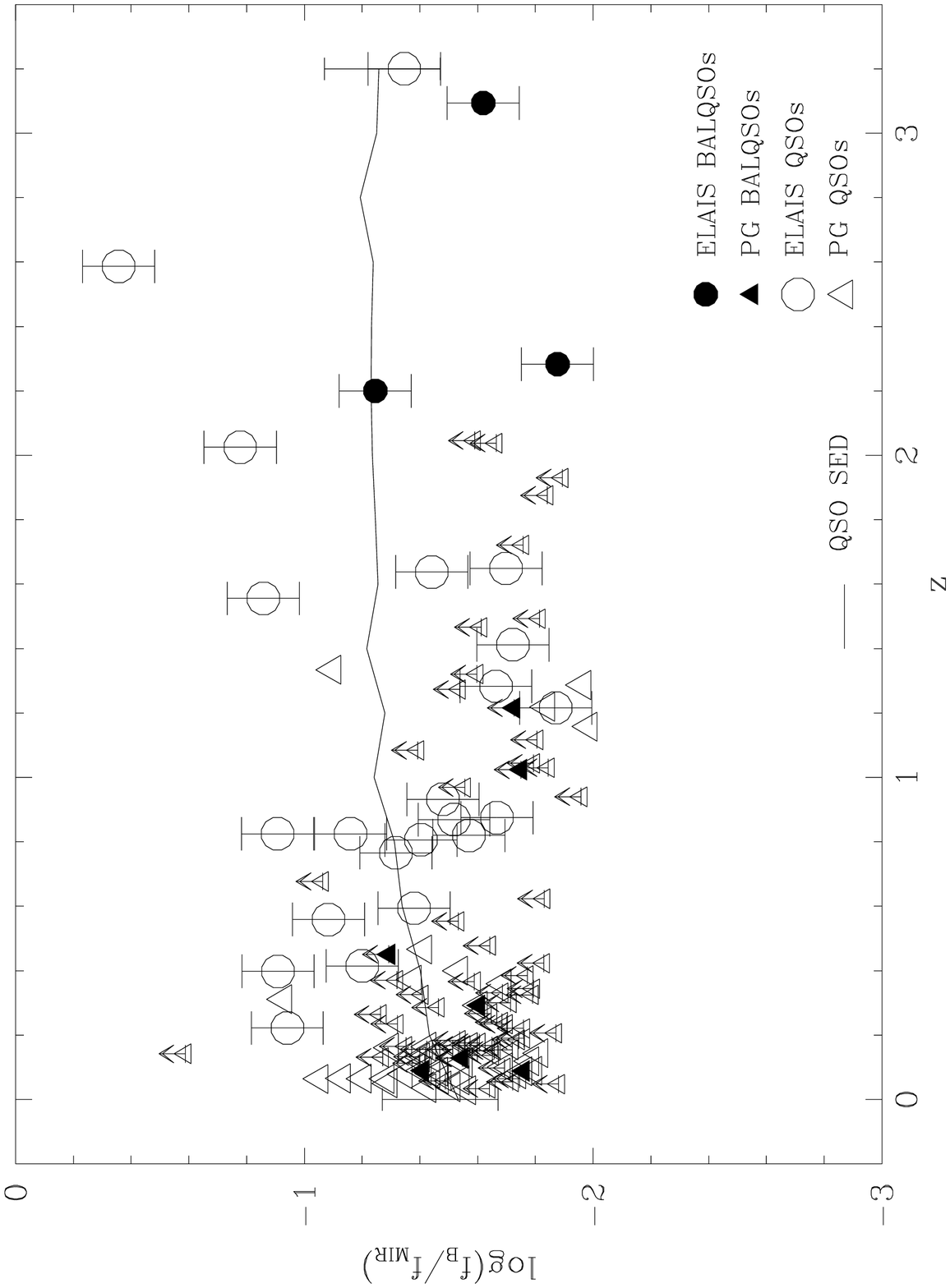}
\end{figure*}


\end{document}